%% file: V2irr-rev01.tex
\documentclass[final,english]{elsarticle}
\usepackage[T1]{fontenc}
\usepackage[latin9]{inputenc}
\usepackage{varioref}
\usepackage{units}
\usepackage{amsmath}
\usepackage{amssymb}
\usepackage{graphicx}
\usepackage{esint}
\usepackage{caption}
\usepackage{subcaption}

\pdfpageheight\paperheight
\pdfpagewidth\paperwidth

\journal{Elsevier}
\usepackage{upgreek}
\usepackage[bookmarks,bookmarksopen,bookmarksdepth=6]{hyperref}
\usepackage{cases}
\usepackage{url}
\usepackage{lineno}

\usepackage{a4wide}

\usepackage{newtxtext,newtxmath,amsmath}
\makeatother

\usepackage{babel}
\begin{document}
\title{Study of the V$_2^0$ state in neutron-irradiated silicon using photon-absorption measurements}

\author[]{E.~Fretwurst}
\author[]{R.~Klanner\corref{cor1}}
\author[]{J.~Schwandt}
\author[]{A.~Vauth}

\cortext[cor1]{Corresponding author. Email address: Robert.Klanner@desy.de,
 Tel. +49~40~8998~2258.}

\address{ Institute for Experimental Physics, University of Hamburg,
 \\Luruper Chaussee 149, 22761, Hamburg, Germany.}



\begin{abstract}

  Pieces of $n$-type silicon with 3.5\,k$\Omega \cdot $cm resistivity have been irradiated by reactor neutrons to 1\,MeV neutron equivalent fluences  of (1, 5 and 10)\,$\times 10^{16}$\,cm$^{-2}$.
  Using light-transmission measurements, the absorption coefficients have been determined for photon energies, $E_\gamma $, between 0.62 and 1.30\,eV for the samples as irradiated and after 15\,min isochronal annealing with temperatures between $80\,^\circ$C and $330\,^\circ$C.
  The radiation-induced absorption coefficient, $\alpha _\mathit{irr}$, has been obtained by subtracting the absorption coefficient for non-irradiated silicon.
  The $E_\gamma $-dependence of $\alpha _\mathit{irr}$ shows a resonance peak, which is ascribed to the neutral divacancy, V$_2^0$, sitting on a background, and
  $\alpha _\mathit{irr} (E_\gamma )$ is fitted by a Breit-Wigner line shape on a parameterized background.
  It is found that at an annealing temperature of 210\,$^\circ $C the V$_2^0$ intensity is reduced by a factor 2, and that at the meV level, the position and the width of the fitted Breit-Wigner do not change with irradiation dose and annealing.

\end{abstract}


\maketitle
 \pagenumbering{arabic}

\section{Introduction}
 \label{sect:Introduction}

 In spite of major efforts for more than 50 years, the understanding of radiation damage in silicon by hadrons remains a big challenge.
 Reasons are the large number of radiation-induced states in the silicon band gap, the formation of defect clusters and the difficulties of their simulation.
 It is well known that Si recoils from the interaction of neutrons and energetic electrons and hadrons produce locally dense cascades of Si vacancies, V, and interstitials, I, with densities as high as $10^{20}$\,cm$^{-3}$\,\cite{Fleming:2007}.
 Inside these clusters the mobile vacancies can agglomerate to multi-vacancy states, in particular to the divacancy, V$_2$, which is immobile at room temperature.
 In \cite{Lint:1972} it has been estimated that there are about 12 V$_2$\,states per cluster, which have typical dimensions of (5\,nm)$^3$.

 The V$_2$ defect is known to come in four charge states: $+,\,0,\, -,\, --$.
 The neutral V$_2^0$ state, which has an energy $E_c - 0.21$\,eV, can trap a hole or an electron resulting in states with  energies $E_v+0.25$\,eV or $E_c - 0.42$\,eV, respectively, where $E_v$ denotes the energy of the valence and $E_c$ of the conduction band.
 The negative V$_2^-$ state can trap electrons, and the resulting state has the energy $E_c - 0.23$\, eV~\cite{Gill:1997}.
 As discussed in \cite{Fan:1959,Cheng:1966,Pajot:2013}, the V$_2^0$ state gives rise to absorption bands at wavelengths of $1.8\,\upmu$m and $3.9\,\upmu$m, and the V$_2^-$ state at $3.6\,\upmu$m.
 From the absence of photocurrent for illumination with $1.8\,\upmu$m light\,\cite{Fan:1959}, it is concluded that the photons induce transitions from the V$_2^0$ ground state to an excited state with energy $E_\mathit{ex}$.
 A qualitative term scheme, deduced from the results of EPR experiments is shown in\,\cite{Cheng:1966}.
 The NIR absorption data of\,\cite{Fan:1959} (Fig.\,2) also show that the overall intensity of the $1.8\,\upmu$m band is approximately independent of temperature.
 However, the mean photon energy, $E_\mathit{ex}$, decreases from about 730\,meV ($\lambda = 1.70\,\upmu$m) at 100\,K to 690\,meV ($\lambda = 1.79\,\upmu$m) at 300\,K, and the full-width-half-maximum, $\Gamma _\mathit{ex}$, increases from 90\,meV to 110\,meV.

 The annealing of the V$_2^0$ in silicon with different doping and impurities has been studied by several groups.
 An example are the results from\,\cite{Cheng:1966} for 20\,minutes isochronal annealing and temperatures up to 330\,$^\circ $C:
 Significant annealing starts around 150\,$^\circ $C, the 50\,\% point is around 240\,$^\circ $C, and complete annealing is reached at 280\,$^\circ $C.
 From the publications it is however not clear if the overall intensity or the maximum of the $1.8\,\upmu$m band is shown, nor if $E_\mathit{ex}$ or $\Gamma _\mathit{ex}$ change during annealing.

 In this work the light transmission for photon energies, $E_\gamma $, between 0.62 and 1.3\,eV has been measured for phosphorous-doped high-ohmic Si irradiated by reactor neutrons to 1\,MeV neutron equivalent fluences $\Phi$ of (1, 5, 10$)\times 10^{16}$\,cm$^{-2}$ and 15 minutes isochronal annealing in $30\,^\circ $C steps up to $T_\mathit{ann} = 330\,^\circ $C.
 From the transmission the absorbance $\alpha (E_\gamma , T_\mathit{ann}, \Phi) $ has been obtained.
 For the radiation-induced absorbance, $\alpha _\mathit{irr}$, the absorption data for non-irradiated Si from \cite{Green:2021} were subtracted.
 The $\alpha _\mathit{irr}$\,spectra for $0.62$\,eV$ \leq E_\gamma \leq 1.05$\,eV were fitted by a Breit-Wigner function for the V$_2^0$ excitation and a phenomenological background parametrisation to determine the values of the V$_2^0$ intensity, $I_{\mathrm{V}_2^0}$, and of the Breit-Wigner parameters $E_\mathit{\mathrm{V}_2^0}$ and $\Gamma _\mathit{\mathrm{V}_2^0}$, as a function of $\Phi $ and $T_\mathit{ann}$.

 The next section gives an overview over the Si samples and the measurement techniques.
 This is followed by a discussion of the analysis method.
 Finally the results are presented and their relevance discussed.

 \section{Samples and light-transmission measurements}
  \label{sect:Samples}

 For the study phosphorous-doped float-zone silicon crystals with approximately 3.5\,k$\Omega \cdot$cm resistivity and 3\,mm thickness were used.
 They were irradiated by reactor neutrons \cite{IJS:Irrad} to 1\,MeV neutron equivalent fluences of $\Phi = (1, 5, 10) \times 10^{16}\,\mathrm{cm}^{-2}$.
 The neutron flux has been $7 \times 10^{12}$\,cm$^{-2} \, $s$^{-1}$, resulting in an irradiation time $t_\mathit{irr} \approx 4$\,h for the fluence of $\Phi = 10^{17}\,\mathrm{cm}^{-2}$.
 The estimated uncertainties of $\Phi $ are 10\,\%.

 The temperature during irradiation, $T_\mathit{irr}$, has not been measured.
 An estimate of $T_\mathit{irr}$ is obtained from the results for the V$_2^0$ absorption band presented in Fig.\,\ref{fig:aV2T} in Sect.\,\ref{sect:Analysis}.
 An Arrhenius model for the time dependence of the annealing of the V$_2^0$ state, $\mathrm{d}\alpha(t)/\mathrm{d}t = \alpha(t) \cdot k_0 \cdot e^{- E_A/(k_B \cdot T)}$, is fitted to the $\Phi = 10^{17}$\,cm$^{-2}$ annealing data of the  $\alpha (T_\mathit{ann})$\,values of the V$_2^0$\,peak for the different isochronal-annealing temperatures, and the frequency factor, $k_0$, and the activation energy, $E_A$, are determined.
 Figure\,\ref{fig:aV2T} shows that $\alpha (\Phi )$ before annealing is not proportional to $\Phi$, which is ascribed to the annealing during the irradiation.
 The ratio of $\alpha(\Phi = 10^{17}\, \mathrm{cm}^{-2})$ to the linear extrapolation from $\alpha(\Phi = 10^{16}\, \mathrm{cm}^{-2})$, is $R = 0.84$.
 Assuming a constant temperature during the irradiation, the Arrhenius model gives $T_\mathit{irr} = -(E_A / k_B)\cdot 1/ \left( \ln \left(-\ln(R)/(k_0 \cdot t_\mathit{irr}) \right) \right)$, with the Boltzmann constant $k_B$.
 From the data a value $T_\mathit{irr} = 70\,^\circ$C is obtained with an estimated uncertainty of about $10\,^\circ$C.
 The samples were not cooled during the transport from Ljubljana to Hamburg, which took about one day.
 The annealing during the transport is estimated to be negligible compared to the annealing during irradiation.
 Apart from the times of irradiation, transport, annealing and measurements, the samples were kept at $-30\,^\circ$C.

\begin{figure}[!ht]
   \centering
   \begin{subfigure}[a]{\textwidth}
    \includegraphics[width=\textwidth]{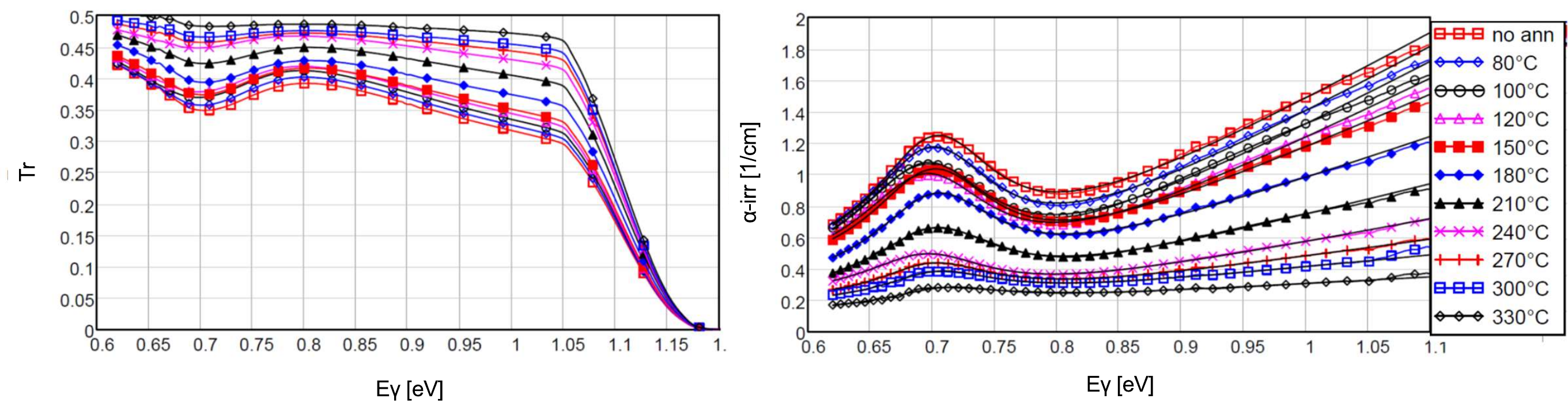}
    \caption{ }
    \label{fig:FigTransE16}
   \end{subfigure}%
  \newline
   \begin{subfigure}[a]{\textwidth}
    \includegraphics[width=\textwidth]{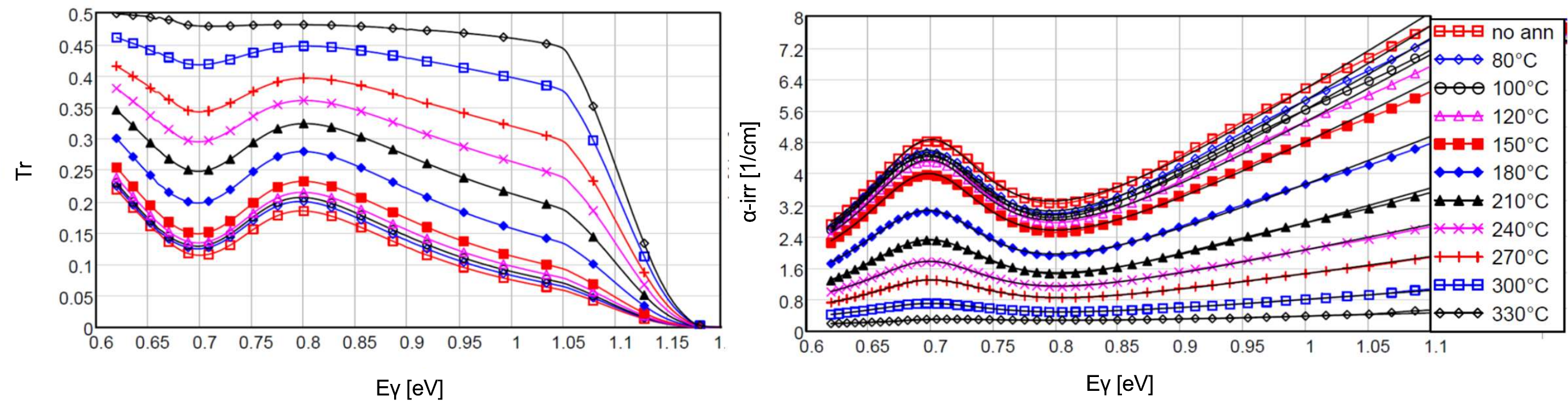}
    \caption{ }
    \label{fig:FigTrans5E16}
   \end{subfigure}%
  \newline
   \begin{subfigure}[a]{\textwidth}
    \includegraphics[width=\textwidth]{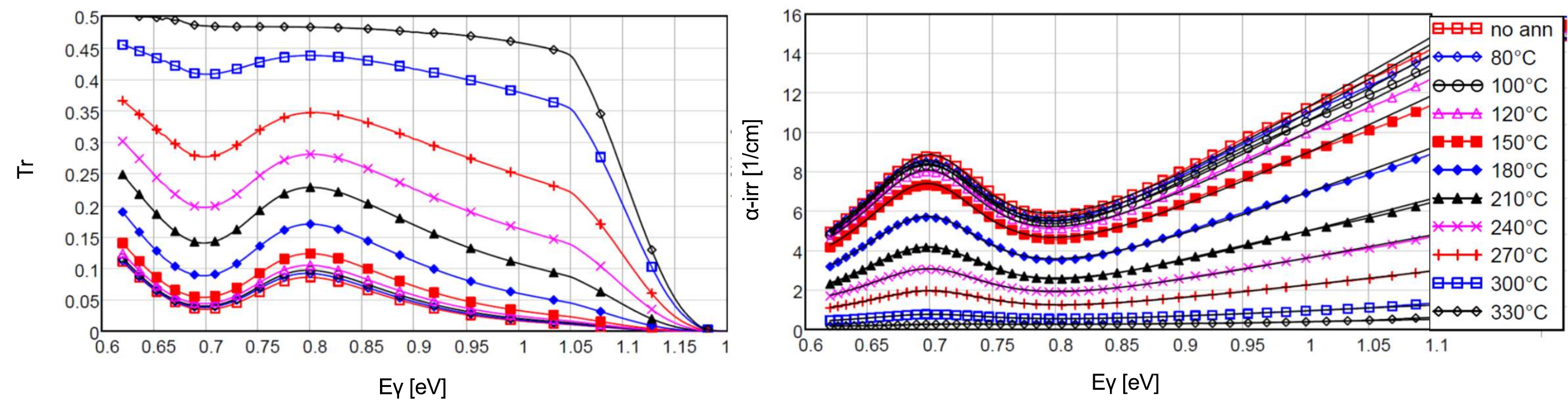}
    \caption{ }
    \label{fig:FigTrans10E16}
   \end{subfigure}%
   \caption{Measured transmission, \emph{Tr}, (left side) and radiation-induced absorbance $\alpha _\mathit{irr}$, (right side) for the different annealing steps as a function of $E_\gamma$ for
   (a) $\Phi = 1 \times 10^{16}$\,cm$^{-2}$,
   (b) $\Phi = 5 \times 10^{16}$\,cm$^{-2}$ and
   (c) $\Phi = 10 \times 10^{16}$\,cm$^{-2}$.
   In this and in other figures markers are used to distinguish the different data sets.
   The distance in $E_\gamma $ of the individual data points is about 1\,meV.
   For $\alpha _\mathit{irr}(E_\gamma$) also the results of the fits for $0.62\, \mathrm{eV} \leq E_\gamma \leq 1.05$\,eV described in the text, are shown as a thin black lines. }
  \label{fig:FigTrans}
 \end{figure}

 For the transmission measurements an Agilent CARY 5000 UV-VIS-NIR \cite{Peest:2017} was used.
 The wavelength range was 0.95 to 2\,$\upmu$m and the temperature about 295\,K.
 The first measurements were performed before annealing.
 For the isochronal annealing the samples were heated for 15 min to temperatures of (80, 100, 120, 150, 180, 210, 240, 270, 300, 330) $^\circ $C.
 The samples were stored at $-30\,^\circ $C, except for the times of the irradiation, transport, annealing and measurement.
 Fig.\,\ref{fig:FigTrans} shows the measured transmission as a function of $E_\gamma $\,[eV] = 1.24$/\lambda \,[\upmu $m].

 The maximum value of \emph{Tr}, which is reached, approximately independent of $\Phi$ at $E_\gamma = 0.62$\,eV for $T_\mathit{ann} = 330\,^\circ$C, is about 0.52.
 The value is close to the expectation of 0.53 from the reflection losses at the two Si-air boundaries of the samples calculated using the Fresnel formulae for normal incidence and the values of the refractive index of Si from \cite{Green:2021}.
 The left side of Fig.\,\ref{fig:FigTrans} shows that \emph{Tr} decreases with increasing $\Phi$, and increases with increasing $T_\mathit{ann}$.
 This is expected, as radiation damage increases with $\Phi$ and decreases with annealing.
 The $E_\gamma $\,dependence of \emph{Tr} has a minimum at $E_\gamma \approx 0.7$\,eV with a full width of $\approx 0.1$\,eV, which is the absorption band of the V$_2^0$\,state.
 For higher $E_\gamma$ values \emph{Tr} decreases slowly up to $\approx 1.05$\,eV, and rapidly above.
 As discussed in \cite{Klanner:2022}, $E_\gamma = 1.040$\,eV is the threshold for the excitation of an electron from the valence to the conduction band by the absorption of an optical transverse phonon.

 \section{Analysis of the absorption data and results}
  \label{sect:Analysis}

 From the measured transmission, $Tr(\lambda)$, the absorption coefficient, $\alpha (\lambda)$, is obtained using the standard formulae \cite{Scharf:2020}:

 \begin{equation}\label{eq:a-data}
  \alpha(\lambda) = \frac{1}{d} \cdot \ln\Bigg(\frac{\mathit{Tra}(\lambda) ^2+\sqrt{\mathit{Tra}(\lambda)^4 + 4 \cdot \mathit{Ref}(\lambda)^2 \cdot \mathit{Tr}(\lambda) ^2} } {2 \cdot \mathit{Tr}(\lambda)} \Bigg)
 \end{equation}
 with $\mathit{Ref}(\lambda) = \big(n(\lambda) - 1 \big)^2 / \big(n(\lambda) + 1 \big)^2 $ and
 $\mathit{Tra}(\lambda) = 1 - \mathit{Ref}(\lambda)$.
 The thickness of the sample is $d$, $\mathit{Ref}(\lambda)$ is the reflection of a single air-silicon interface, $\mathit{Tra}(\lambda)$ the corresponding transmission, and $n(\lambda)$ the index of refraction of silicon taken from \cite{Green:2021}.

 When using Eq.\ref{eq:a-data} two important assumptions are made:
 (1) The refractive index of silicon does not change with irradiation, and
 (2) the transmission is dominated by absorption and the effects of scattering can be ignored.

 To the authors' knowledge no investigation of the $\Phi $\,dependence of the refractive index has been reported in the literature.
 An estimate can be made using the data presented in \cite{Klanner:2022a}:
 The measured geometrical capacitance of irradiated pad diodes, which is reached either at high bias voltages or high frequencies, changes by less than 0.2\,\% for $\Phi$ between $3 \times 10^{15}$ and $13 \times 10^{15}$\,cm$^{-2}$.
 From this observation it can be concluded that $\Delta \varepsilon_\mathit{Si}/\varepsilon_\mathit{Si} < 0.2\,\%$, and with $n = \sqrt{\varepsilon_\mathit{Si}}$, that \emph{n} changes by less than 0.1\,\% up to $\Phi = 13 \times 10^{15}$\,cm$^{-2}$.
 Also for the $\Phi $\,dependence of the scattering no measurements are available.
 The effect is that the scattered photons miss the sensor of the spectral-photometer and therefore reduce the measured transmission.
 Thus, the absorbance $\alpha $ includes the contribution from scattering.

 The following analysis uses $\alpha _\mathit{irr} (\Phi) = \alpha (\Phi) - \alpha (\Phi = 0)$.
 For the absorbance of non-irradiated Si, $\alpha (\Phi = 0)$, the data from \cite{Green:2021} are used.
 These data only extend to wavelengths $\lambda = 1.45\,\upmu$m, where $\alpha = 5 \times 10^{-8}$\,cm$^{-1}$ corresponding to an attenuation length of 20\,km, which can not be measured with a 0.3\,cm Si slab.
 For $\lambda > 1.45\,\upmu$m the constant value of $5 \times 10^{-8}$\,cm$^{-1}$ is assumed.
 It is noted that the subtraction of $\alpha (\Phi = 0)$ has no effect on the V$_2^0$ absorption band, which only covers the region $\lambda \gtrsim 1.6\,\upmu$m.
 However, the subtraction of $\alpha (\Phi = 0)$ allows extending the fits to $\alpha $ to lower $\lambda $-, and thus to higher $E_\gamma $-values.

 The right-hand side of Fig.\,\ref{eq:a-data} shows $\alpha _\mathit{irr} (E_\gamma)$ for the different annealing temperatures, $T_\mathit{ann}$, and the three $\Phi $\,values.
 It can be seen that $\alpha _\mathit{irr}$ increases with $\Phi $, and decreases with $T_\mathit{ann}$.
 At $E_\gamma \approx 0.7$\,eV, the centre of the V$_2^0$\,absorption band, a clear peak with a Breit-Wigner line shape is observed, whose amplitude decreases with $T_\mathit{ann}$ and which is practically zero for $T_\mathit{ann} = 330\,^\circ$C.
 It is also observed that $\alpha _\mathit{irr}$ outside and under the V$_2^0$\,absorption band shows a similar decrease.

 To quantitatively extract the parameters of the V$_2^0$-line shape and the background the following parametrisation is used:
 \begin{equation}\label{eq:a-model}
   \alpha (E_\gamma) = I_{\mathrm{V}_2^0} \cdot \left( \mathrm{BW}(E_\gamma ;E_{\mathrm{V}_2^0},\Gamma _{\mathrm{V}_2^0})+ \sqrt{a^2 + \left( b \cdot \mathrm{max} (0,E_\gamma - E_{BG}) \right)^2} \right),
 \end{equation}
 where BW is the is the normalized Breit-Wigner probability density with mean $E_{\mathrm{V}_2^0}$ and width $\Gamma _{\mathrm{V}_2^0}$, and
 $I_{\mathrm{V}_2^0}$ is the contribution of the V$_2^0$\,band to $\alpha _\mathit{irr}$.
 The background shape is parameterized by the constant $a$ for $E_\gamma $ below the background-transition energy, $E_\mathit{BG}$, and the square root of the sum of  $a^2$ and the square of a straight line with slope $b$ for $E_\gamma > E_\mathit{BG}$.
 Both, the Breit-Wigner and the background shape are multiplied with $I_{\mathrm{V}_2^0}$.
 In this way the background parameters are constant if the signal-to-background ratio is constant.
 Several ad-hoc parameterizations of the background have been tried, but Eq.\,\ref{eq:a-model} gave the best description of the data.
 However, the results for the parameters of the Breit-Wigner function are hardly influenced by the background parametrizations which are compatible with the data.

 Eq.\,\ref{eq:a-model} is fitted to the $\alpha _\mathit{irr}$\,data on the right-hand side of Fig.\,\ref{fig:FigTrans} in the range $0.62\,\mathrm{eV} \leq E_\gamma \leq 1.05$\,eV.
 The fit results can be seen as black continuous lines.
 As can be seen, the data are well described by the fit, with an \emph{rms} deviation of about $5 \times 10^{-4}$ of the mean $\alpha _\mathit{irr}$\,value of the individual data sets, which comprise 819 $E_\gamma$ values each.

  \begin{figure}[!ht]
   \centering
   \begin{subfigure}[a]{0.5\textwidth}
    \includegraphics[width=\textwidth]{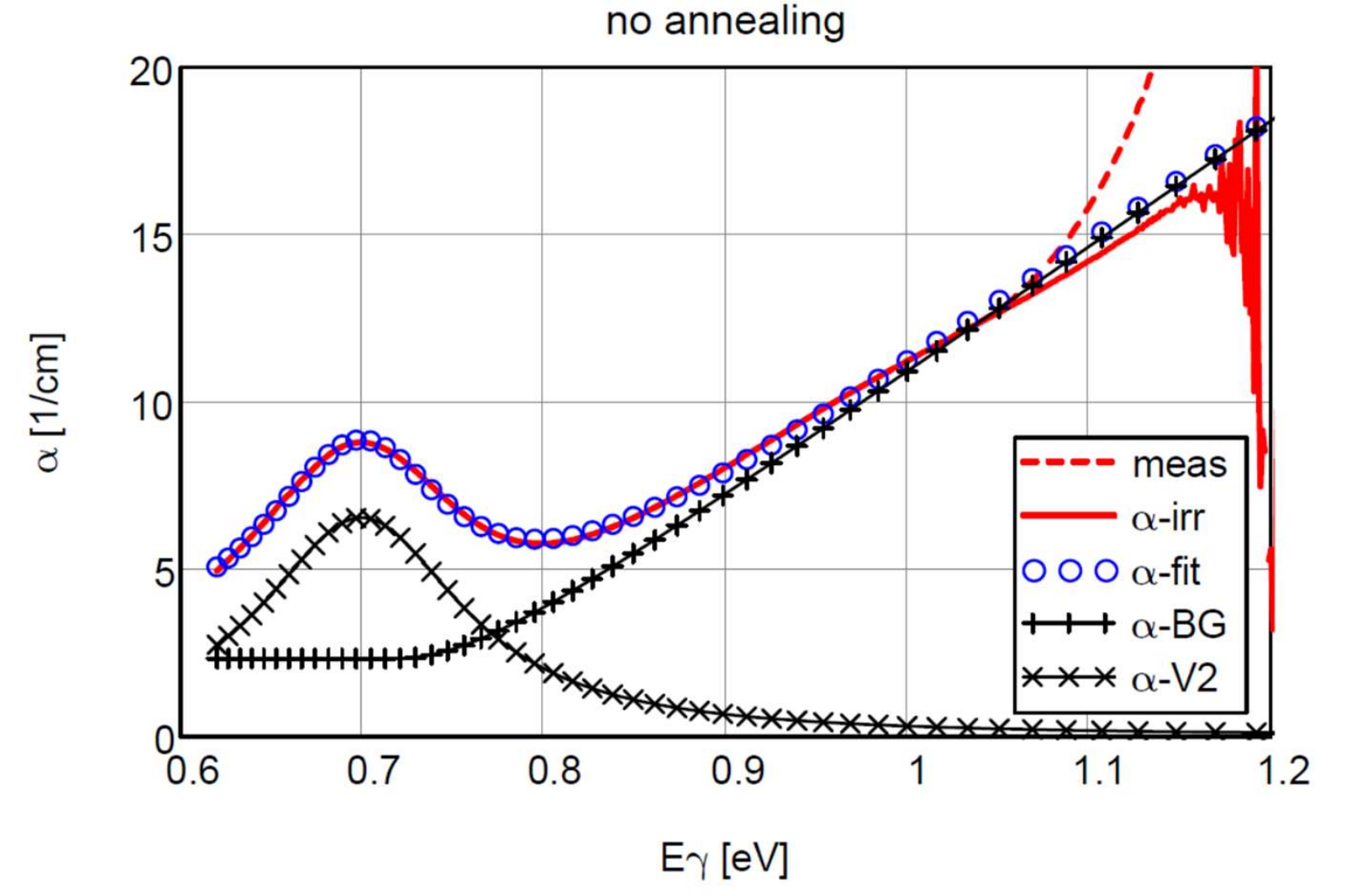}
    \caption{ }
    \label{fig:alphaE17T20}
   \end{subfigure}%
    ~
   \begin{subfigure}[a]{0.5\textwidth}
    \includegraphics[width=\textwidth]{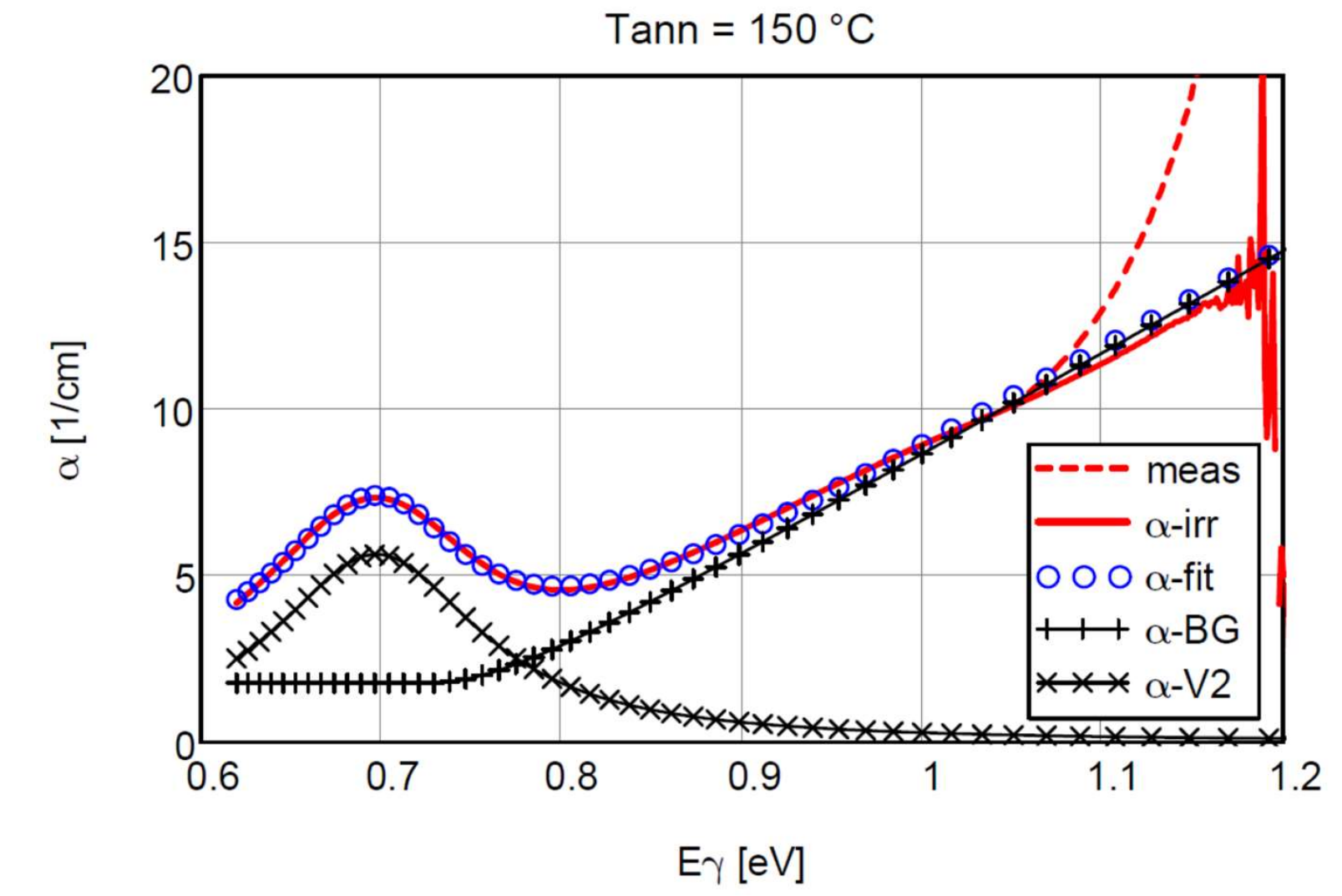}
    \caption{ }
    \label{fig:alphaE17T150}
   \end{subfigure}%
  \newline
   \begin{subfigure}[a]{0.5\textwidth}
    \includegraphics[width=\textwidth]{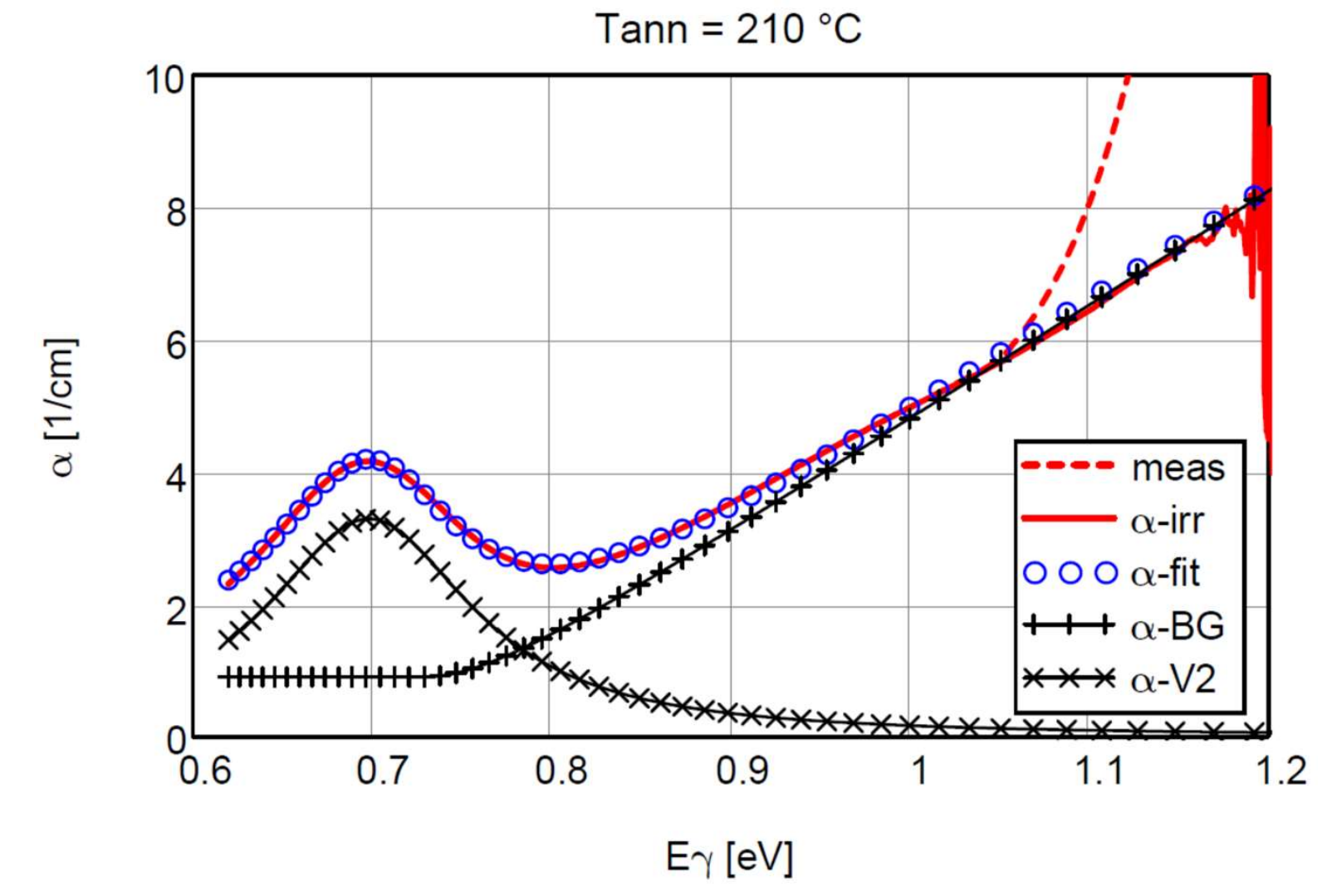}
    \caption{ }
    \label{fig:alphaE17T20}
   \end{subfigure}%
    ~
   \begin{subfigure}[a]{0.5\textwidth}
    \includegraphics[width=\textwidth]{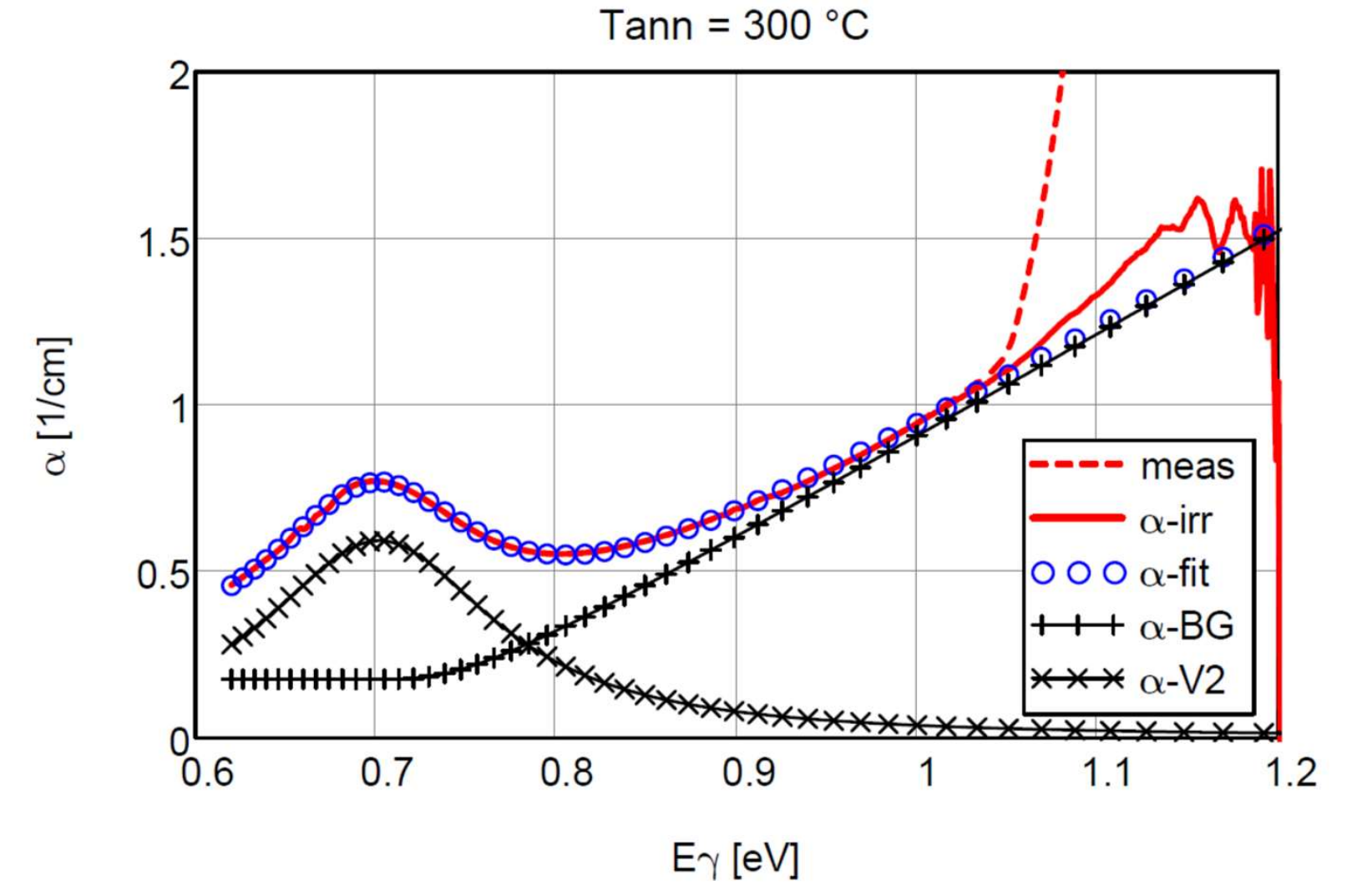}
    \caption{ }
    \label{fig:alphaE17T150}
   \end{subfigure}%
   \caption{Measured $\alpha(E_\gamma )$, $\alpha_\mathit{irr}(E_\gamma )$ and $\alpha_\mathit{fit}(E_\gamma )$ for the data of $\Phi = 10^{17}$\,cm$^{-2}$ for
   (a) no annealing,
   (b) $T_\mathit{ann} = 150\,^\circ$C,
   (c) $T_\mathit{ann} = 210\,^\circ$C, and
   (d) $T_\mathit{ann} = 300\,^\circ$C.
   In addition, the absorbance for the V$_2^0$ and the background, as determined by the fit in the range $0.62\,\mathrm{eV} \leq E_\gamma \leq 1.05$\,eV described in the text, are shown.
   The symbols are markers and the spacing between the individual measurements is about 1\,meV.
   The dashed lines represent the $\alpha(E_\gamma )$\,values determined from the transmission data. }
  \label{fig:alphaE17}
 \end{figure}

 More details of the fits can be seen in Fig.\,\ref{fig:alphaE17} which shows the results for the $\Phi = 10^{17}$\,cm$^{-2}$ data before annealing, and for $T_\mathit{ann} = 150, 210$ and $ 300\,^\circ$C.
  The V$_2^0$ absorption band, which peaks around 0.7\,eV, is followed by an approximately linear increase up to about 1.05\,eV, where $\alpha $ rapidly increases because the photons can excite electrons from the valence to the conduction band.
 Above $E_\gamma \approx 1.15$\,eV the transmission is less than $10^{-4}$ and the measurements are dominated by noise.
 For $\alpha_\mathit{irr}$, where the contribution of $\alpha $ for the non-irradiated silicon is subtracted, the fast rise above 1.05\,eV is essentially absent.
 The open circles are the fits of Eq.\,\ref{eq:a-model} to $\alpha_\mathit{irr}(E_ \gamma)$, and the inclined and vertical crosses show the contributions of the V$_2^0$~Breit-Wigner and of the background, respectively.
 It can be seen that the parameters of the V$_2^0$ do not change with $T_\mathit{ann}$, except for the intensity, which decreases by more than an order of magnitude.
 The results for the $\Phi = 1$ and $5 \times 10^{16}$\,cm$^{-2}$ data are similar, only the V$_2^0$\,intensities are smaller.

  \begin{figure}[!ht]
   \centering
   \begin{subfigure}[a]{\textwidth}
    \includegraphics[width=\textwidth]{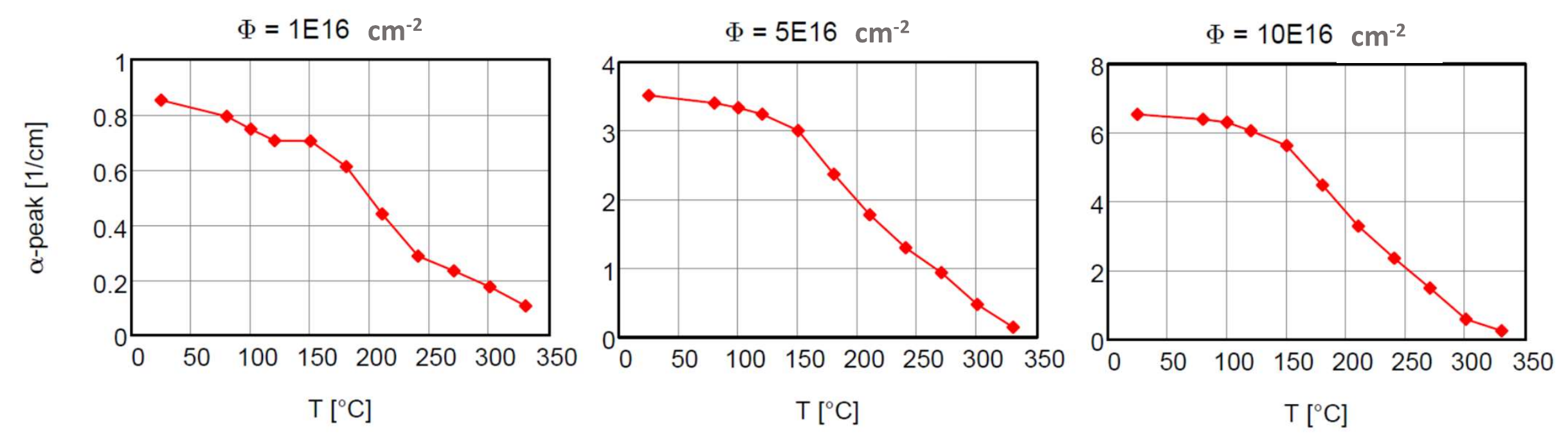}
    \caption{ }
    \label{fig:aV2T}
   \end{subfigure}%
  \newline
   \begin{subfigure}[a]{\textwidth}
    \includegraphics[width=\textwidth]{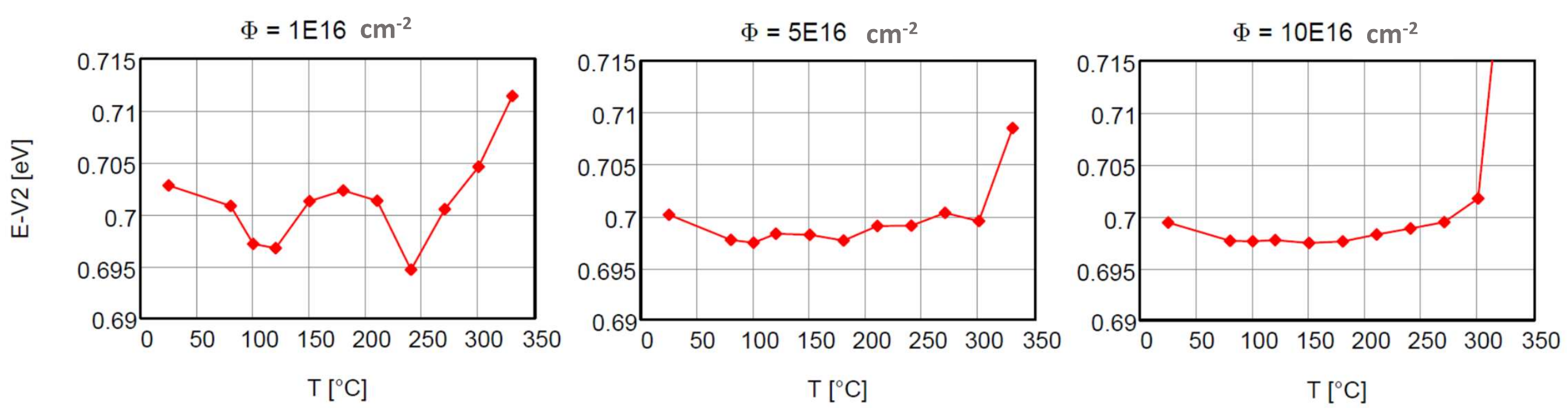}
    \caption{ }
    \label{fig:EV2T}
   \end{subfigure}%
  \newline
   \begin{subfigure}[a]{\textwidth}
    \includegraphics[width=\textwidth]{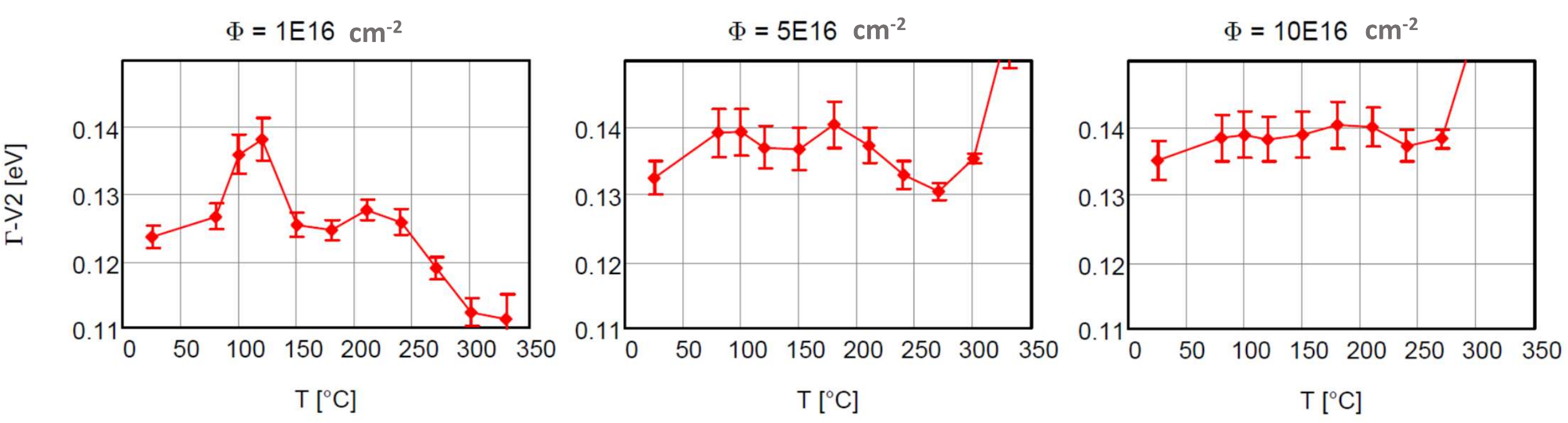}
    \caption{ }
    \label{fig:GV2T}
   \end{subfigure}%
   \caption{Parameters of the V$_2^0$\,absorption band as a function of $T_\mathit{ann}$ and $\Phi$ from the fit of Eq.\,\ref{eq:a-model} to $\alpha _\mathit{irr} (E_\gamma)$ for $0.62\,\mathrm{eV} \leq E_\gamma \leq 1.05$\,eV.
   (a) Amplitude of the Breit-Wigner, $\alpha $-\emph{peak} in cm$^{-1}$,
   (b) mean energy of the Breit-Wigner, $E _{\mathrm{V}_2^0}$, in eV, and
   (c) full width of the Breit-Wigner, $\Gamma _{\mathrm{V}_2^0}$, in eV.
 }
  \label{fig:V2T}
 \end{figure}

 Figure\,\ref{fig:V2T} shows the Breit-Wigner parameters as a function of $T_\mathit{ann}$ for the three $\Phi $\,values investigated from the fits of Eq.\,\ref{eq:a-model} to the $\alpha _\mathit{irr} (E_\gamma)$\,data for $0.62\,\mathrm{eV} \leq E_\gamma \leq 1.05$\,eV.
 From Fig.\,\ref{fig:aV2T}, which shows the amplitude of the Breit-Wigner $\alpha $-\emph{peak}, it is seen that the $\alpha $-\emph{peak} before annealing (shown in the plot at $T_\mathit{ann} = 20\,^\circ$C) increases with $\Phi $.
 As a function of $T_\mathit{ann}$ the V$_2^0$\,state anneals.
 As expected, the annealing curves for the different $\Phi $\,values are similar.
 At about $210\,^\circ$C, the amplitude is reduced by a factor 2.
 The annealing curves are very similar to results reported in the literature\,\cite{Cheng:1966}.

 Figure\,\ref{fig:EV2T} shows the central energy of the Breit-Wigner peak, $E _{\mathrm{V}_2^0}$, as a function of $T_\mathit{ann}$.
 For $T_\mathit{ann}$\,values up to $300\,^\circ$C the $\Phi = 1 \times 10^{16}$\,cm$^{-2}$ data show fluctuations of $\pm 4$\,meV, whereas for the higher $\Phi $\,values the fluctuations are about $\pm 1$\,meV.
 The larger fluctuations are ascribed to weak V$_2^0$\,signals for low $\Phi $ and for high $T_\mathit{ann}$.
 The value of $699 \pm 1$\,meV is compatible with the value at room temperature which can be extracted from Fig.\,2 of Ref.\,\cite{Fan:1959}.
 It is concluded that the position of the Breit-Wigner does not change within $\pm 1$\,meV.

 Figure\,\ref{fig:GV2T} shows the full width of the Breit-Wigner peak, $ \Gamma _{\mathrm{V}_2^0}$, as a function of $T_\mathit{ann}$.
 Here the fluctuations are bigger than for $E _{\mathrm{V}_2^0}$ because of the strong anti-correlation between background and $\Gamma _{\mathrm{V}_2^0}$.
 The value found for the two high-$\Phi $\,data sets and $T_\mathit{ann} < 300\,^\circ$C is $138 \pm 2$\,meV, without any trend of a $T_\mathit{ann} $\,dependence.
 Again, this value is compatible with the value at room temperature which can be extracted from Fig.\,2 of Ref.\,\cite{Fan:1959}.

  \begin{figure}[!ht]
   \centering
   \begin{subfigure}[a]{\textwidth}
    \includegraphics[width=\textwidth]{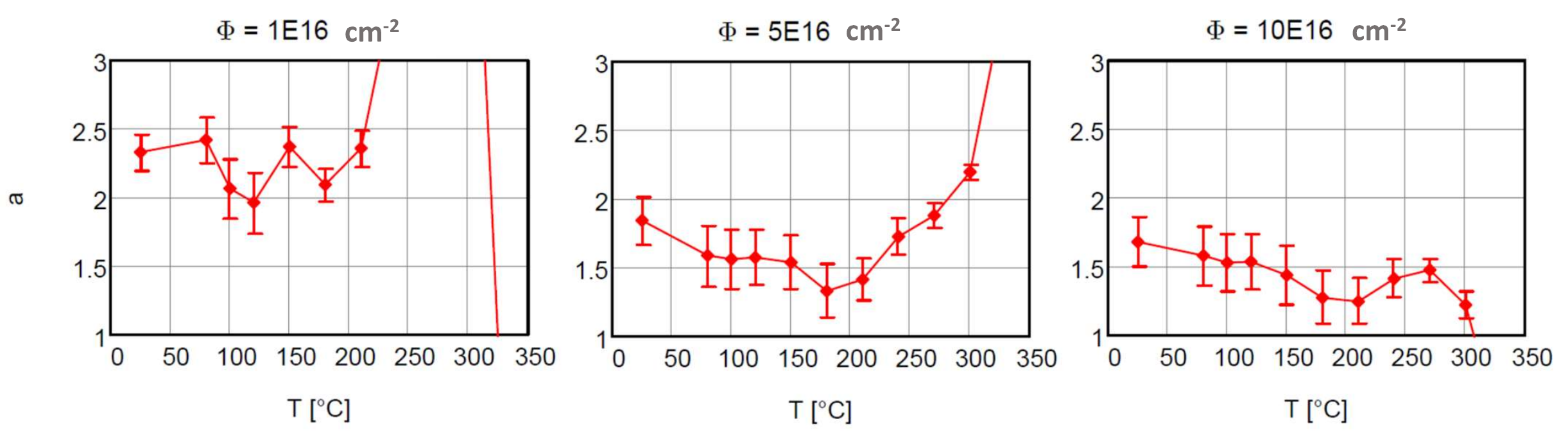}
    \caption{ }
    \label{fig:aBGT}
   \end{subfigure}%
  \newline
   \begin{subfigure}[a]{\textwidth}
    \includegraphics[width=\textwidth]{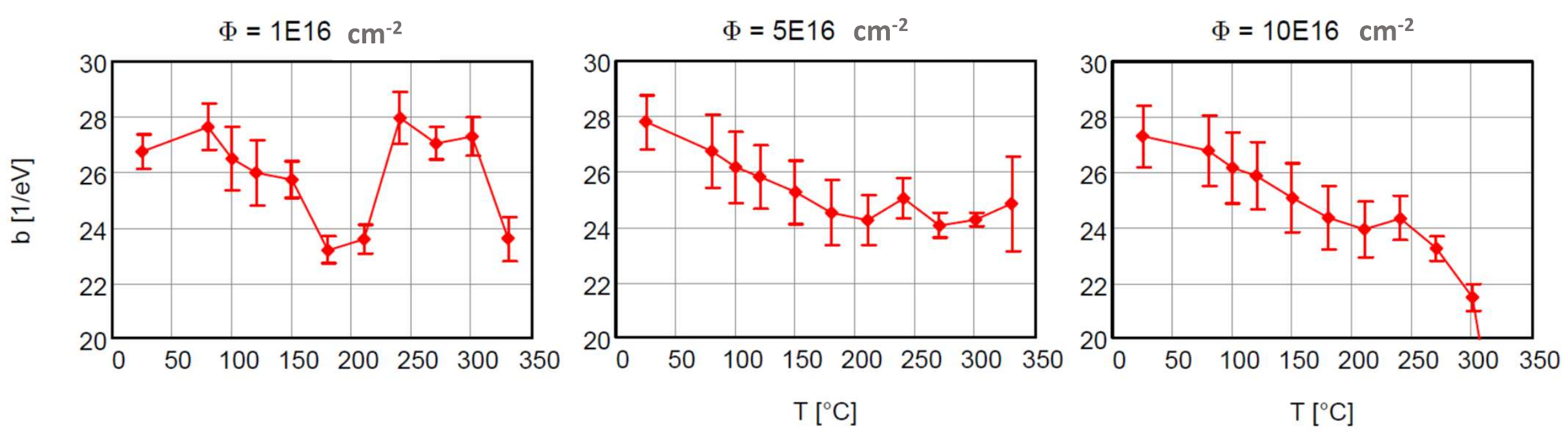}
    \caption{ }
    \label{fig:bBGT}
   \end{subfigure}%
  \newline
   \begin{subfigure}[a]{\textwidth}
    \includegraphics[width=\textwidth]{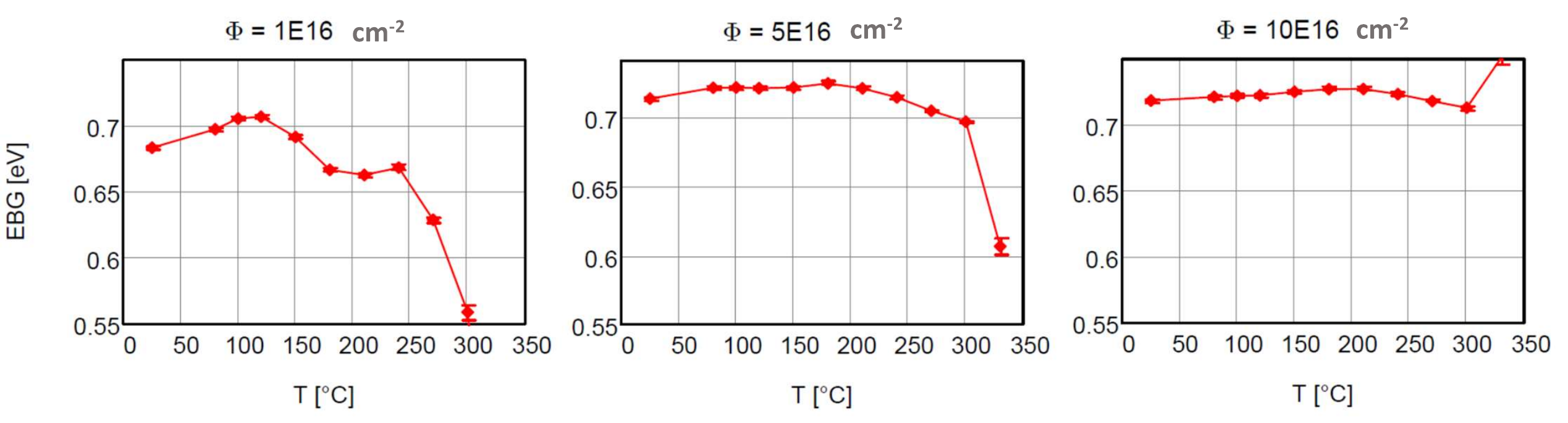}
    \caption{ }
    \label{fig:EBGT}
   \end{subfigure}%
   \caption{Parameters of the background shape as a function of $T_\mathit{ann}$ and $\Phi$ from the fit of Eq.\,\ref{eq:a-model} to $\alpha _\mathit{irr} (E_\gamma)$ for $0.62\,\mathrm{eV} \leq E_\gamma \leq 1.05$\,eV.
   (a) Constant term, \emph{a},
   (b) slope, \emph{b}, and
   (c) background-transition energy, $E_{BG}$.
   All parameters are relative to the intensity of the V$_2^0$\,band, and thus describe the shape of the background.
 }
  \label{fig:BGT}
 \end{figure}

 In Fig.\,\ref{fig:BGT} the shape parameters of the background are presented as a function of $T_\mathit{ann}$ for the three  $\Phi$\,values.
  It can be seen that, in particular the values for $\Phi = 1 \times 10^{16}$\,cm$^{-2}$ show large fluctuations.
  Figure\,\ref{fig:aBGT}, shows that as a function of $T_\mathit{ann}$ the value of \emph{a}, which is the ratio of the background at the V$_2^0$\,peak to the V$_2^0$\,intensity, is approximately constant.
 For the slope \emph{b}, which is shown in Fig.\,\ref{fig:bBGT}, a decrease from about 27\,eV$^{-1}$ to about 23\,eV$^{-1}$ is observed.
 The background-transition energy, where a value $E_{BG} \approx 725$\,meV is determined,  does not depend on $T_\mathit{ann}$.
 It can be concluded that at the position of the V$_2^0$\,peak the annealing of the background is the same as the annealing of the V$_2^0$.
 Above the V$_2^0$\,peak the decrease of \emph{b} with $T_\mathit{ann}$ indicates that the annealing of the background is stronger than the V$_2^0$ annealing.


 The main results of this section are:
 The NIR transmission, measured at room temperature, of silicon irradiated by reactor neutrons has been used to study the excitation of the V$_2^0$ state by photons for 15-minutes isochronal annealing for temperatures up to 330\,$^\circ$C.
 It is found that at an annealing temperature of about 210\,$^\circ$C the V$_2^0$\,density is reduced by a factor 2, and that the energy and full width of the V$_2^0$\,absorption band do not change with annealing temperature.
 It is also observed, that the annealing of the background under the V$_2^0$ band is similar to the annealing of the V$_2^0$.

 \section{Comments to the results}
  \label{sect:Results}

 The results on the V$_2^0$\,absorption and annealing reported in this paper are certainly not new, and similar results have been reported since the late 1950-ies.
 What is new, are the fits to the radiation-induced absorption coefficient, $\alpha _\mathit{irr} (E_\gamma)$, allowing to precisely determine the intensity, energy and width of the V$_2^0$\,absorption band and its background.
 It is found, that for reactor-neutron fluences between (1 and 10)\,$\times 10^{16}$\,cm$^{-2}$ and subsequent isochronal annealing up to a temperature of 330\,$^\circ $C, the values of the energy and the width of the V$_2^0$\,absorption band do not change at the 1\,meV level, when the V$_2^0$\,intensity decreases by more than a factor 10 because of annealing.
 As for neutron irradiation the V$_2^0$\,states are known to be mainly located in damage clusters, one can conclude that the cluster environment does not influence the V$_2^0$-excitation spectrum.
 A probably related observation is reported in \cite{Klanner:2022}, where it is found that the silicon band-gap energy
 changes by less than 1\,meV for neutron irradiation up to fluences of $10^{17}$\,cm$^{-2}$.
 In Ref.\,\cite{Fan:1959} it is shown that the V$_2^0$\,absorption band changes with temperature, which shows that the V$_2^0$-excitation spectrum can be influenced by the environment.
 It is also noted that in DLTS-\,\cite{Svensson:1991} and TSC-\,\cite{Donegani:2018}\,measurements a reduction of the electron-emission time constants of radiation-induced states in clusters, compared to the emission-time constants for isolated defects, is reported.

  \section{Summary}
  \label{sect:Conclusions}

  Slabs of $n$-type silicon with about 3.5\,k$\Omega \cdot $cm resistivity and a thickness of 3\,mm have been irradiated by reactor neutrons to fluences of (1, 5, 10)$\times 10^{16}$\,cm$^{-2}$.
  The transmission for light with wavelengths between 0.9 and 2.0\,$\upmu$m has been measured at room temperature of the samples, as irradiated and after 15\,min isochronal annealing with temperatures between $80\,^\circ$C and $330\,^\circ$C.
  From the transmission, the absorption coefficient $\alpha (E_\gamma )$ has been determined.
  By subtracting literature values of $\alpha (E_\gamma )$ for non-irradiated silicon, $\alpha _\mathit{irr}(E_\gamma )$, the radiation-induced contribution to $\alpha (E _\gamma)$, has been obtained.
  The $E_\gamma $\,dependence of $\alpha _\mathit{irr}(E_\gamma )$ shows a peak for $E_\gamma \approx 700 $\,meV on a background.
  The $\alpha _\mathit{irr} (E _\gamma)$\,spectrum has been fitted by a Breit-Wigner resonance curve, which is ascribed to the excitation of the neutral divacancy, V$_2^0$, on a parameterized absorption background.
  The 6 parameters of the model have been determined as a function of irradiation fluence and annealing temperature.

 The main results of the paper are:
 \begin{itemize}
   \item The model, which has 6 parameters, 3 for the V$_2^0$\,absorption band and 3 for the background, precisely describes $\alpha _\mathit{irr} (E_\gamma )$ for 0.62\,eV$\, \le E_\gamma \le$\,1.05\,eV.
   \item The energy and the width of the V$_2^0$\,absorption band agree with literature values.
   \item The intensity of the V$_2^0$\,absorption before annealing deviates from a proportionality to the fluence, which is ascribed to annealing during the irradiation.
   \item The energy and the width of the V$_2^0$\,absorption band do not depend on radiation fluence and annealing.
   \item At an annealing temperature of 210\,$^\circ$C the V$_2^0$\,absorption is reduced by a factor of approximately 2.
   \item The annealing of the background at the V$_2^0$\,peak is similar to the annealing of the  V$_2^0$\,states. For photon energies above the V$_2^0$\,peak, the annealing of the background is somewhat stronger.
 \end{itemize}

 The quantitative results of the study can be used to check models of the V$_2^0$ complex in hadron-irradiated silicon.
 In addition, they can be used to correct for the radiation-induced change of photon absorption in experiments which characterise radiation-damaged silicon detectors using NIR light.


\section*{Acknowledgements}
 \label{sect:Acknowledgement}
  We thank Lukas Terkowski from the group of Roman Schnabel of the Institute of Laser Physics of Hamburg University for his help with the photo-spectrometer measurements, and Vladimir Cindro and Gregor Kramberger for the neutron irradiations performed at the TRIGA reactor of the Jozef Stefan Institute, Ljubljana.
  We are also thankful to
  Ioana Pintilie for information on the properties of radiation-induced defects.
  The  work was partially supported by the Deutsche Forschungsgemeinschaft (DFG, German Research Foundation) under Germany's Excellence Strategy -- EXC 2121 Quantum Universe - 390833306.

 \input{bibliography}

  \label{sect:Bibliography}



\end{document}

%% file: bibliography.tex
%

\section{List of References}

%% file: V2irr-rev01.bbl
\begin{thebibliography}{0}

 \bibitem{Fleming:2007}
  R.M.\,Fleming et al.,
   \emph{Effects of clustering on the properties of defets in neutron irradiated silicon},
    Journal of Applied Physics 102 (2007) 043711.

 \bibitem{Lint:1972}
  V.A.J.\,van Lint, R.E.\,Leadon and J.F.\,Colwell,
   \emph{Energy dependence of displacement effects in semiconductors},
    IEEE Transactions on Nuclear Science 19 (1972) 181--185.

 \bibitem{Gill:1997}
  K.\,Gill, G.\,Hall and B.\,MacEvoy,
   \emph{Bulk damage effects in irradiated silicon detectors due to clustered divacancies},
    Journal of Applied Physics 82 (1997) 126--136.

 \bibitem{Fan:1959}
  H.Y.\,Fan and A.K.\,Ramdas,
   \emph{Infrared absorption and photoconductivity in irradiated silicon},
    Journal of Applied Physics 30 (1959) 1127--1134.

 \bibitem{Cheng:1966}
  L.J.\,Cheng, J.C.\,Corelli, J.W.\,Corbett and G.D.\,Watkins,
   \emph{1.8-, 3.3- and 3.9-$\upmu$ bands in irradiated silicon: Correlation with the divacancy},
    Physical Review 152 (1966) 761--774.

 \bibitem{Pajot:2013}
  B.\,Pajot and B.\,Clerjaud,
   \emph{Optical Absorption of Impurities and Defects in Semiconducting Crystals},
    Springer Series in Solid-State Sciences 169, Springer Verlag Berlin, Heidelberg 2013.

  \bibitem{Green:2021}
   M.A.~Green,
   \emph{Improved silicon optical parameters at 25\,$^\circ $C, 295\,K and 300\,K including temperature coefficients},
     Progress in Photovoltaics 30 (2021) 164--179.


 \bibitem{IJS:Irrad}
  K.~Ambrozic, G.~Zerovnik and L.~Snoj,
   \emph{Computational analysis of the dose rates at JSI TRIGA reactor irradiation facilities},
    Applied Radiation and Isotopes, 130 (2017) 140--152.


 \bibitem{Peest:2017}
  C.~Peest et al.,
   \emph{Instrumentation-related uncertainty of reflectance and transmission measurements with a two-channel spectrophotometer},
    Review of Scientific Instruments, 88 (2017) 015105.


 \bibitem{Klanner:2022}
  R.~Klanner, S.~Martens, J.~Schwandt and A.\,Vauth,
   \emph{Study of the band-gap energy of radiation-damaged silicon},
    New Journal of Physics 24 (2022) 073017.

  \bibitem{Scharf:2020}
    C.~Scharf, F.~Feindt and R.~Klanner,
     \emph{Influence of radiation damage on the absorption of near-infrared light in silicon},
      Nuclear Instruments and Methods in Physics Research A 968 (2020) 163955.

  \bibitem{Klanner:2022a}
    R.~Klanner and J.\,Schwandt,
     \emph{Can the electric field in radiation-damaged silicon pad diodes be determined by admittance and current measurements?},
      Nuclear Instruments and Methods in Physics Research A 1028 (2022) 166360.

 \bibitem{Svensson:1991}
  B.G.~Svensson et al.,
   \emph{Divacancy acceptor levels in ion-irradiated silicon},
    Physical Review B 43 (1991) 2292--2298.

  \bibitem{Donegani:2018}
  E.~Donegani et al.,
   \emph{Study of point- and cluster-defects in radiation-damaged silicon},
   Nuclear Instruments and Methods in Physics Research A 898 (2018) 15--23.


\end{thebibliography}
